\documentclass[aps,prd,twocolumn,10pt,nofootinbib,longbibliography]{revtex4-1}
\usepackage{amssymb}
\usepackage{graphicx}
\usepackage{amsmath}
\usepackage{hyperref}
\usepackage{subfigure}
\usepackage{multirow}
\usepackage{setspace}
\usepackage{verbatim}
\usepackage{float}
\usepackage{color}
\usepackage{ulem}
\usepackage[utf8]{inputenc}

\begin{document}

\title{Bubble expansion at strong coupling}

\author{Li Li$^{1,2,3,4}$}
\email{liliphy@itp.ac.cn}

\author{Shao-Jiang Wang$^{1}$}
\email{schwang@itp.ac.cn (Corresponding author)}

\author{Zi-Yan Yuwen$^{1,2}$}
\email{yuwenziyan@itp.ac.cn (Corresponding author)}

\affiliation{$^1$CAS Key Laboratory of Theoretical Physics, Institute of Theoretical Physics, Chinese Academy of Sciences, Beijing 100190, China}
\affiliation{$^2$School of Physical Sciences, University of Chinese Academy of Sciences (UCAS), Beijing 100049, China}
\affiliation{$^3$School of Fundamental Physics and Mathematical Sciences, Hangzhou Institute for Advanced Study (HIAS), University of Chinese Academy of Sciences (UCAS), Hangzhou 310024, China}
\affiliation{$^4$Peng Huanwu Collaborative Center for Research and Education, Beihang University, Beijing 100191, China.}

\begin{abstract}
The cosmological first-order phase transition (FOPT) can be of strong dynamics but with its bubble wall velocity difficult to be determined due to lack of detailed collision terms. Recent holographic numerical simulations of strongly coupled theories with a FOPT prefer a relatively small wall velocity linearly correlated with the phase pressure difference between false and true vacua for a planar wall. In this Letter, we have analytically revealed the non-relativistic limit of a planar/cylindrical/spherical wall expansion of a bubble strongly interacting with the thermal plasma. The planar-wall result reproduces the linear relation found previously in the holographic numerical simulations. The results for cylindrical and spherical walls can be directly tested in future numerical simulations. Once confirmed, the bubble wall velocity for a strongly coupled FOPT can be expressed purely in terms of the hydrodynamics without invoking the underlying microphysics.
\end{abstract}
\maketitle

\section{Introduction}

The first-order phase transition (FOPT) is ubiquitous in nature from the gas-liquid PT in daily life to the counterpart~\cite{Mazumdar:2018dfl,Hindmarsh:2020hop,Caldwell:2022qsj} in the early Universe, the latter of which plays an indispensable role in probing new physics~\cite{Cai:2017cbj,Bian:2021ini} via the associated stochastic gravitational-wave (GW) backgrounds~\cite{Caprini:2015zlo,Caprini:2019egz} as well as the induced large curvature perturbations~\cite{Liu:2022lvz} or even the produced primordial black holes~\cite{Liu:2021svg}. Despite the success of modeling GW spectra from numerical simulations, a key parameter usually left undetermined is the terminal wall velocity $\xi_w$ for a steady-state bubble expansion, before which the initially nucleated static bubble starts to accelerate under a driving fore that could be eventually balanced by a backreaction force~\cite{Wang:2022txy,Wang:2023kux}. Whether the bubbles had reached the terminal wall velocity when they largely collide with each other crucially determines whether the GWs are dominated by wall collisions or fluid motions~\cite{Cai:2020djd,Lewicki:2022pdb} .

Previous attention~\cite{Bodeker:2009qy,Bodeker:2017cim,BarrosoMancha:2020fay,Hoche:2020ysm,Ai:2021kak,Vanvlasselaer:2020niz,Balaji:2020yrx,Gouttenoire:2021kjv,Dorsch:2021nje,DeCurtis:2022hlx} has much focused on the relativistic limit of the wall velocity, $\gamma_w=1/\sqrt{1-\xi_w^2}\gtrsim\mathcal{O}(1)$, and the corresponding backreaction-force dependence on $\gamma_w$ to see whether the bubble wall could run away without bound and if not, the precise computation of the terminal wall velocity from fully solving the Boltzmann equations~\cite{Moore:1995ua,Moore:1995si,Konstandin:2014zta,Laurent:2020gpg,Laurent:2022jrs}. Nevertheless, FOPTs in the early Universe can be of strong-coupling dynamics like composite dark sectors~\cite{Schwaller:2015tja}. However, for a strongly coupled FOPT, it is barely feasible to write down the exact collision terms. Furthermore, the usual perturbative field theoretic calculation of the vacuum decay rate from the effective potential could also break down at strong coupling, which motivates recent computations of an effective action from holography \cite{Ares:2021ntv,Ares:2021nap,Morgante:2022zvc} but still with the terminal wall velocity left unspecified.


Remarkably, it was recently found in holographic numerical simulations~\cite{Bea:2021zsu,Janik:2022wsx} of strongly coupled systems that there is a novel linear correlation~\cite{Bigazzi:2021ucw} between the phase pressure difference of false and true vacua and the terminal velocity of the planar wall, though a similar relation has not been reported yet for the cylindrical wall~\cite{Bea:2022mfb}, let alone the more realistic but most probabilistic case of a spherical wall with $O(3)$ symmetry for a thermal FOPT~\cite{Coleman:1977py,Linde:1980tt,Linde:1981zj}. Furthermore, the holographic numerical simulations also suggest a non-relativistic terminal wall velocity. This could be understood as a bubble wall strongly interacting with the thermal plasma, the backreaction force is so rapidly growing that it only takes a very short moment of time for acceleration before the backreaction force had already balanced the driving force. Unlike the relativistic case where the local thermal equilibrium around the bubble wall cannot be sustained as the particles had not had enough time to fully thermalize before the bubble wall swept over, the non-relativistic case could largely retain the local thermal equilibrium and hence the perfect fluid hydrodynamic approximation near (but not right at) the wall interface. This is consistent with a recent observation~\cite{Baggioli:2021tzr} that the hydrodynamics could still be valid even far from equilibrium (see also the ``unreasonable effectiveness"~\cite{hong2020,Noronha-Hostler:2015wft} of hydrodynamics out of equilibrium).

In this \textit{Letter}, we will adopt hydrodynamics~\cite{Espinosa:2010hh,Leitao:2010yw,Wang:2022txy} to analytically derive the non-relativistic limit of bubble expansion at strong coupling for all different wall geometries with planar, cylindrical, and spherical symmetries, not only reproducing the pre-mentioned linear relation for the planar wall but also predicting new relations with logarithmic and quadratic dependences for the cylindrical and spherical walls that can be directly tested in future  numerical simulations in particularly in holography. Once confirmed, these relations can be used to express the bubble wall velocity purely in terms of the  hydrodynamics without turning to the underlying microphysics.

\section{Bubble expansion hydrodynamics}

Assuming a thin-wall steady-state bubble in a flat background of thermal plasma without shear and bulk viscosity, the total energy-momentum tensor of a scalar-plasma system can be cast into a perfec-fluid form $T^{\mu\nu}=(e+p)u^\mu u^\nu+p\eta^{\mu\nu}$ as a wall-fluid system~\cite{Wang:2022txy}, where the total energy density and pressure read $e=e_f+e_\phi=e_f+V_0$ and $p=p_f+p_\phi=p_f-V_0$ with $V_0$ the zero-temperature part of the total effective potential $V_\mathrm{eff}(\phi,T)=V_0(\phi)+V_T(\phi,T)$ that admits false and true vacua at $\phi_+$ and $\phi_-$, respectively. Note that the effective potential in the finite-temperature field theory as the free energy density is the opposite of the total pressure, $V_\mathrm{eff}\equiv\mathcal{F}\equiv-p$, then the fluid part of the pressure simply reads $p_f=p-p_\phi=-V_\mathrm{eff}+V_0=-V_T$. The four-velocity $u^\mu\equiv\mathrm{d}x^\mu/\mathrm{d}\tau$ of bulk fluids can be defined in the planar, cylindrical, and spherical coordinate systems with $x^\mu=(t,z,x=0,y=0)$, $x^\mu=(t,\rho,\varphi=0,z=0)$, and $x^\mu=(t,r,\theta=0,\varphi=0)$, respectively, and yields $u^\mu=\gamma(v)(1,v,0,0)$ in the background plasma frame with the three-velocity $v\equiv\mathrm{d}x^1/\mathrm{d}x^0$ and corresponding Lorentz factor $\gamma(v)\equiv1/\sqrt{1-v^2}$. If the initial size of the critical bubble can be neglected, then the bubble expansion is self-similar with a single self-similar coordinate $\xi\equiv x^1/x^0$. In particular, the steady-state wall position $x_w^1(x^0)=\xi_w t$ is fixed at the terminal wall velocity $\xi_w$ in the self-similar coordinate. In the local frame moving with $\xi_w$ along the $x^1$-direction, the wall-frame bulk fluid velocity reads $u^\mu=\bar{\gamma}(1,-\bar{v},0,0)$ with $-\bar{v}=(v-\xi_w)/(1-v \xi_w)\equiv-\mu(\xi_w, v)$ and $\bar{\gamma}\equiv\gamma(\bar{v})=1/\sqrt{1-\bar{v}^2}$, where the minus sign in front of $\bar{v}$ is introduced to ensure a positive $\bar{v}$ for later convenience, and the abbreviation $\mu(\zeta, v(\xi))\equiv(\zeta-v)/(1-\zeta v)$ is introduced for the Lorentz transformation of the bulk fluid velocity $v(\xi)$ in the plasma frame into a local frame comoving with a velocity $\zeta$.

Equation of motions (EOMs) of bulk fluids are obtained from projecting the conservation equation of total energy-momentum tensor, $\nabla_\mu T^{\mu\nu}=0$, parallel along and perpendicular to the bulk fluid flow direction~\cite{Espinosa:2010hh} as,
\begin{align}
D\frac{v}{\xi} & = \gamma(v)^2(1-\xi v)\left( \frac{\mu(\xi,v)^2}{c_s^2} - 1 \right) \frac{\mathrm{d}v}{\mathrm{d}\xi}, \label{eq:fluidEOM1}\\
\frac{\mathrm{d}\ln w}{\mathrm{d}\xi}&=\gamma(v)^2\mu(\xi,v)\left(\frac{1}{c_s^2}+1\right)\frac{\mathrm{d}v}{\mathrm{d}\xi},\label{eq:fluidEOM2}
\end{align}
where $D = 0$, $1$, and $2$ correspond to planar, cylindrical, and spherical walls, respectively \cite{Leitao:2010yw}, $c_s=\sqrt{\partial_\xi p/\partial_\xi e}$ is the sound speed, and $w=e+p$ is the enthalpy. The first EOM~\eqref{eq:fluidEOM1} can be numerically solved with junction conditions across the bubble wall and shock front (if any),
\begin{align}
w_-\bar{v}_-\bar{\gamma}_-^2&=w_+\bar{v}_+\bar{\gamma}_+^2,\label{eq:junctionwall}\\
w_L\tilde{v}_L\tilde{\gamma}_L^2&=w_R\tilde{v}_R\tilde{\gamma}_R^2,\label{eq:junctionshock}
\end{align}
where $w_\pm$, $\bar{v}_\pm$, and $\bar{\gamma}_\pm\equiv\gamma(\bar{v}_\pm)$ are the enthalpy, wall-frame fluid velocity, and corresponding Lorentz factors just right in front and back of the wall, respectively, while $w_{L/R}$, $\tilde{v}_{L/R}$, and $\tilde{\gamma}_{L/R}\equiv\gamma(\tilde{v}_{L/R})$ are the enthalpy, shock-front-frame fluid velocity, and corresponding Lorentz factors just right in back and front of the shock front, respectively. With a solved fluid profile $v(\xi)$, the enthalpy profile $w(\xi)$ is simply obtained by integrating the second EOM~\eqref{eq:fluidEOM2} from some known  $w(\xi_0)$.

\section{Strongly coupled FOPT}

For a strongly coupled FOPT, the EOS could be well approximated in the MIT bag model~\cite{Chodos:1974je} with the sound velocity $c_s=1/\sqrt{3}$ by collecting the energy density $e_\pm=a_\pm T_\pm^{1+c_s^{-2}}+V_0^\pm$ and pressure $p_\pm=c_s^2 a_\pm T_\pm^{1+c_s^{-2}}-V_0^\pm$ from the vacuum energy and ideal gas, where $V_0^\pm\equiv V_0(\phi_\pm)$ and $a_\pm\equiv(\pi^2/30)g_\mathrm{eff}^\pm$ are the vacuum potential energy density and effective number of relativistic degrees of freedom in the false and true vacua, respectively, and $T_\pm$ are the plasma temperatures just right in front and back of the wall, respectively. With a bag EOS, the wall-frame fluid velocities near the wall can be related by~\cite{Espinosa:2010hh}
\begin{align}\label{eq:vpmbar}
\bar{v}_+=\frac{1}{1+\alpha_+}\left(X_+\pm\sqrt{X_+^2-(1+\alpha_+)(c_s^2-\alpha_+)}\right)
\end{align}
with $X_\pm\equiv\bar{v}_-/2\pm c_s^2/(2\bar{v}_-)$, where $\alpha_+\equiv(1+c_s^2)\Delta V_0/w_+$ is the strength factor of the FOPT characterizing the released vacuum energy density difference $\Delta V_0\equiv V_0^+-V_0^-$ with respect to the radiation energy density $a_+T_+^4\equiv3w_+/4$ just right in front of the wall. Similarly, we can define an asymptotic strength factor $\alpha_N\equiv(1+c_s^2)\Delta V_0/w_N$ from $w_N=(1+c_s^2)a_+T_N^{1+c_s^{-2}}$ with the subscript ``$N$'' for the value at null infinity $\xi=1$ hereafter. Note that $\alpha_+w_+=\alpha_Nw_N=(1+c_s^2)\Delta V_0$. The fluid motion of detonation expansion ($c_s<\bar{v}_-<\bar{v}_+\equiv\xi_w$) picks the plus-sign branch of \eqref{eq:vpmbar}, while the deflagration expansion ($\bar{v}_+<\bar{v}_-=\xi_w<c_s$) and hybrid expansion ($\bar{v}_+<\bar{v}_-=c_s$) picks the minus-sign branch of \eqref{eq:vpmbar}.

For the acceleration stage of spherical wall expansion, the effective EOM for the wall position $r_w(t)$ obeys~\cite{Cai:2020djd}
\begin{align}
\left(\sigma+\frac{r_w}{3}\frac{\mathrm{d}p_\mathrm{br}}{\mathrm{d}\gamma_w}\right)\frac{\mathrm{d}\gamma_w}{\mathrm{d}r_w}+\frac{2\sigma\gamma_w}{r_w}=p_\mathrm{dr}-p_\mathrm{br},
\end{align}
where the Lorentz factor $\gamma_w(\dot{r}_w(t))\equiv1/\sqrt{1-\dot{r}_w^2}$ has been re-expressed as a function of $r_w(t)$, $\sigma$ is the bubble wall tension, $p_\mathrm{dr}\equiv \Delta V_\mathrm{eff}\equiv V_\mathrm{eff}(\phi_+)-V_\mathrm{eff}(\phi_-)$ is the driving force (per unit area), and the backreaction force $p_\mathrm{br}\equiv  p_\mathrm{th}+p_\mathrm{fr}$ consists of the thermal force $p_\mathrm{th}$ and friction force $p_\mathrm{fr}$~\cite{Wang:2022txy} due to temperature inhomogeneity and non-equilibrium effects at the vicinity of the wall interface, respectively . We can rearrange the $\gamma_w$-independent and $\gamma_w$-dependent parts of the backreaction force as the LO and non-LO contributions, $p_\mathrm{br}\equiv p_\mathrm{LO}+f(\gamma_w) p_{N\mathrm{LO}}$, with a general function $f(\gamma_w)$. The general solution reads
\begin{align}
\frac{f(\gamma_w)-f(1)}{f(\gamma_w^\mathrm{eq})-f(1)}+\frac{3\gamma_w}{2r_w}=1+\frac{1}{2r_w^3},
\end{align}
where, at the late-time limit $r_w\to\infty$, the Lorentz factor $\gamma_w$ approaches its terminal value $\gamma_w^\mathrm{eq}$ determined by the balance $f(\gamma_w^\mathrm{eq})=(p_\mathrm{dr}-p_\mathrm{LO})/p_{N\mathrm{LO}}$.

To see whether a non-relativistic $\gamma_w^\mathrm{eq}\approx1$ is preferred at strong coupling limit, we can first assume a relativistic $\gamma_w^\mathrm{eq}\gtrsim\mathcal{O}(1)$ for a heuristic proof by contradiction as given below. Under the relativistic limit, the LO contribution~\cite{Bodeker:2009qy} $p_\mathrm{LO}=\Delta m^2T^2/24$ with $\Delta m^2\equiv\sum_ic_ig_i\Delta m_i^2$ sums over all particles ($c_i=1$ for bosons and $c_i=1/2$ for fermions) due to their mass-square changes $\Delta m_i^2\equiv m_i^2(\phi_-)-m_i^2(\phi_+)$. Thus, $p_\mathrm{dr}-p_\mathrm{LO}=\Delta V_0-\frac14\Delta w+\mathcal{O}(m_i^3/T^3)$ is largely independent of gauge couplings for a bag EOS, while the non-LO contribution $f(\gamma_w)p_{N\mathrm{LO}}$ is still under debate~\cite{Bodeker:2009qy,Bodeker:2017cim,Hoche:2020ysm,Gouttenoire:2021kjv}. As a simple illustration, in the relativistic regime, the NLO contribution~\cite{Bodeker:2017cim} $\gamma p_\mathrm{NLO}=\gamma g^2\Delta m_VT^3$ with $g^2\Delta m_V=\sum_ig_i\lambda_i^2\Delta m_i$ sums over only gauge bosons with gauge couplings $\lambda_i$ and mass changes $\Delta m_i\equiv m_i(\phi_-)-m_i(\phi_+)$. Hence, at strong gauge coupling limit $\lambda_i\gg1$, we arrive at a contradiction with our assumption $\gamma_w^\mathrm{eq}\gtrsim\mathcal{O}(1)$, that is
\begin{align}
\gamma_w^\mathrm{eq}=\frac{\Delta V_0-\frac14\Delta w}{T^3\sum\limits_ig_i\lambda_i^2\Delta m_i}\ll1.
\end{align}
Therefore, the terminal wall velocity for a strongly coupled FOPT should be modestly non-relativistic, $\gamma_w^\mathrm{eq}\approx1$.  This is also exactly what was preferably seen in recent holographic numerical simulations~\cite{Bea:2021zsu,Janik:2022wsx}. More rigorous proof requires computing the friction force at the non-relativistic limit reserved for future work.

\section{Pressure difference at strong coupling}

\begin{figure}
\includegraphics[width=0.45\textwidth]{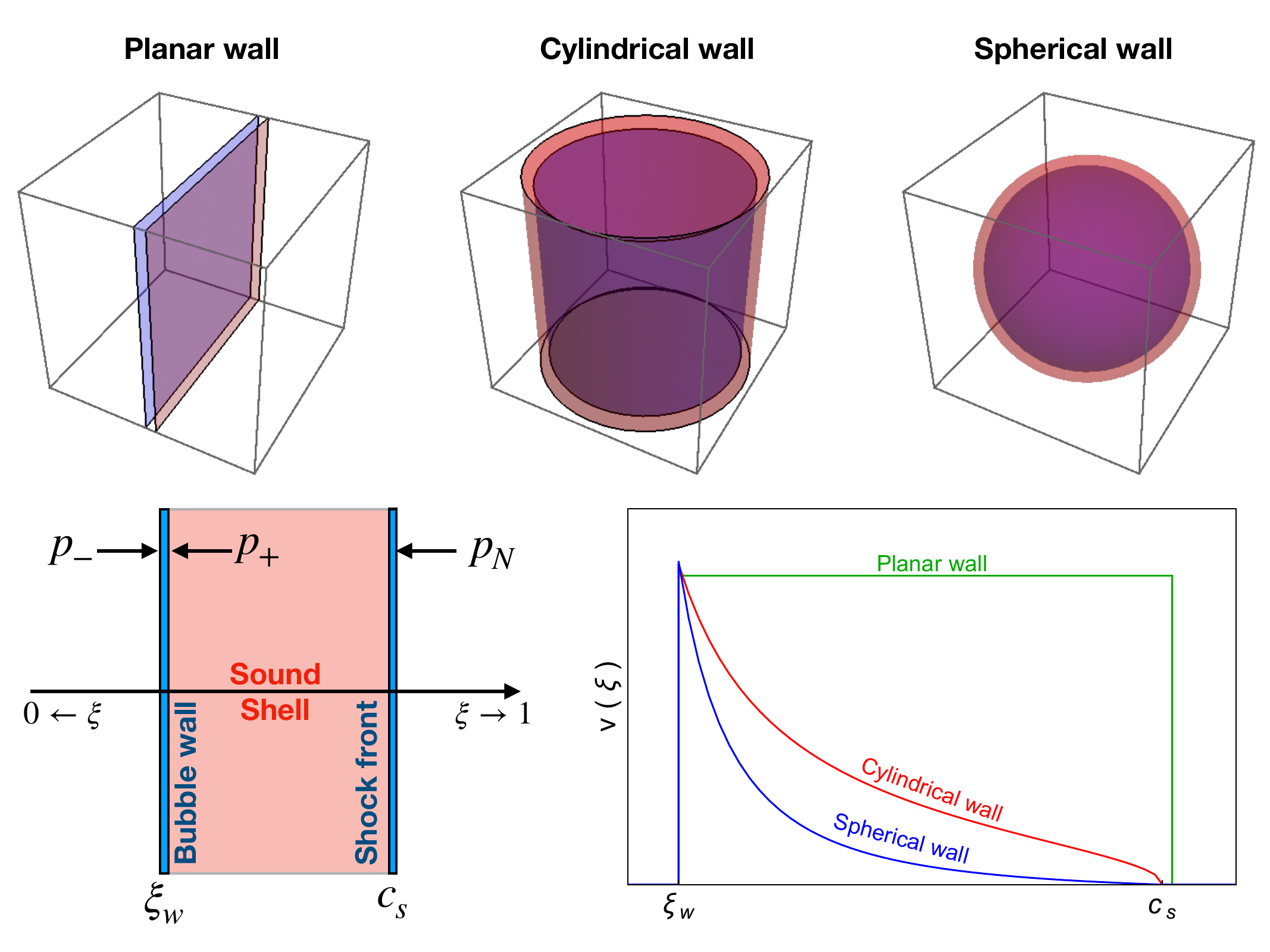}\\
\caption{The schematic illustration for different walls of planar, cylindrical, and spherical geometries (upper row) with corresponding fluid velocity profiles (bottom right) as well as the illustration (bottom left) for the pressure differences $p_\mathrm{dr}=p_--p_N$ and $\Delta_\mathrm{wall}p=p_+-p_-$ acting on the sound shell and bubble wall, respectively. }\label{fig:schematic}
\end{figure}

After the wall approaches a non-relativistic terminal velocity, the prefect-fluid hydrodynamics takes over the subsequent evolution of the deflagration expansion, where a compressive shockwave with the shock front at $\xi_{sh}$ acts as a sound shell in front of the wall. To evaluate the phase pressure difference between the false and true vacua, $p_\mathrm{dr}=\Delta V_\mathrm{eff}=-\Delta p=\Delta\left(-c_s^2aT^4+V_0\right)=-[c_s^2/(1+c_s^2)]\Delta w+\alpha_Nw_N/(1+c_s^2)$ with $\Delta w\equiv w_N-w_O$, we can first equal the enthalpy $w_O$ at the origin to the enthalpy $w_-$ just right behind the wall for deflagration expansion, and then adopt the junction condition~\eqref{eq:junctionwall} at the wall interface with $\bar{v}_+=\mu(\xi_w,v_+)$ and $\bar{v}_-=\xi_w$ to express $w_-$ in terms of $\xi_w$, $v_+$, and $w_+$. Next, $w_+$ can be further expressed in terms of $\xi_w$, $v_+$, and observable parameters at null infinity like $\alpha_N$ and $w_N$ by
\begin{align}
w_+=\frac{(1-v_+^2)\alpha_Nw_N\xi_w}{v_+[c_s^2+(1-c_s^2)v_+\xi_w-\xi_w^2]},
\end{align}
which is equivalent to the minus-sign branch of \eqref{eq:vpmbar}. Finally, the phase pressure difference in units of $w_N$ reads
\begin{align}\label{eq:FdriveDeflagration}
\frac{p_\mathrm{dr}}{w_N}=-\frac{c_s^2}{1+c_s^2}+\frac{c_s^2+v_+^2-(1+c_s^2)v_+\xi_w}{v_+[c_s^2+(1-c_s^2)v_+\xi_w-\xi_w^2]}\cdot\frac{\xi_w\alpha_N}{1+c_s^2}.
\end{align}
Since the minus-sign branch of~\eqref{eq:vpmbar} has expressed  $v_+=\mu(\xi_w,\bar{v}_+(\xi_w,\alpha_+))$ in terms of $\xi_w$ and $\alpha_+$, we only need to find a relation between  $\alpha_+$ and $\alpha_N$, which can be achieved approximately to the leading order in $\xi_w$ for planar, cylindrical, and spherical walls as detailed below. See also Fig.~\ref{fig:schematic} for a schematic illustration.

\subsection{Planar wall}

\begin{figure}
\centering
\includegraphics[width=0.42\textwidth]{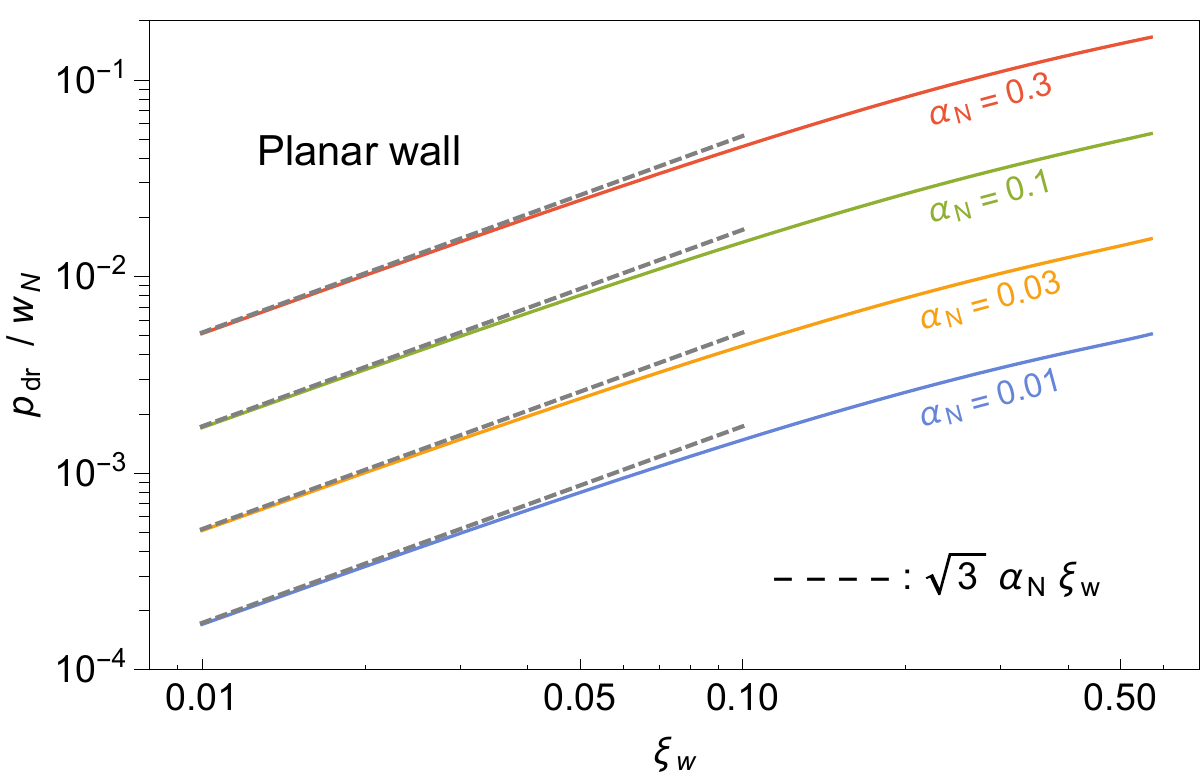}\\
\includegraphics[width=0.42\textwidth]{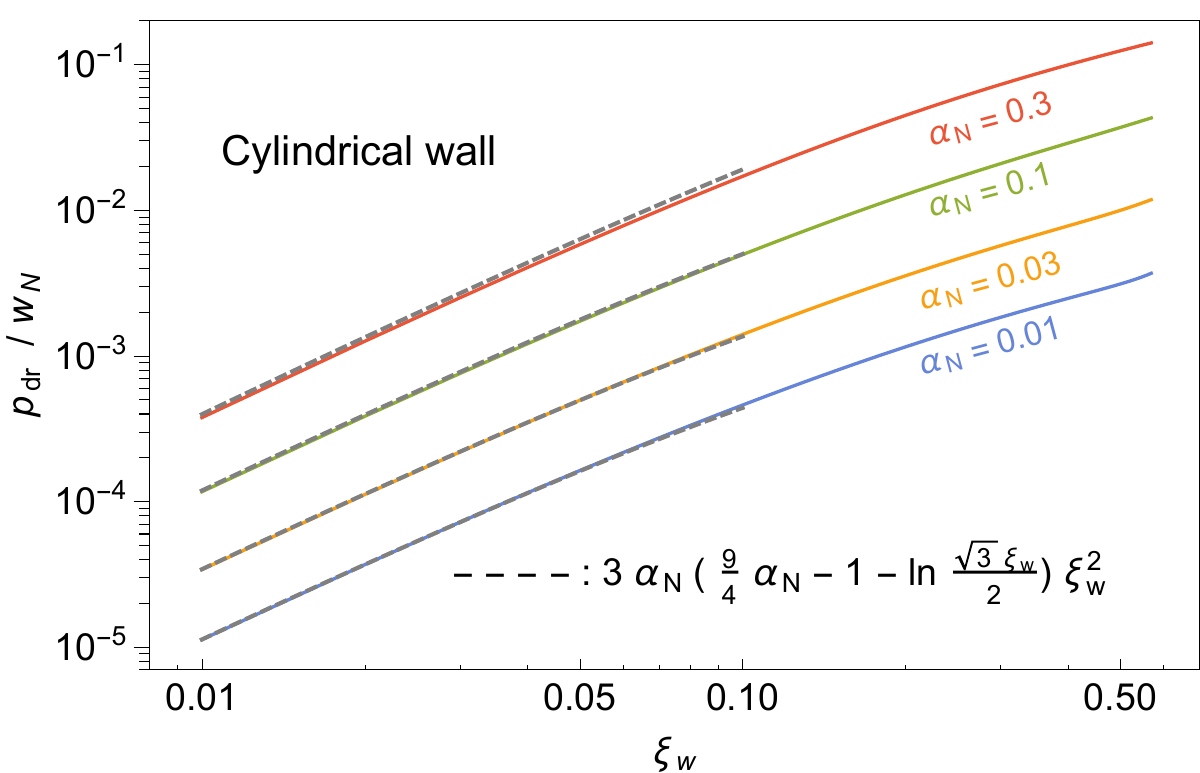}\\
\includegraphics[width=0.42\textwidth]{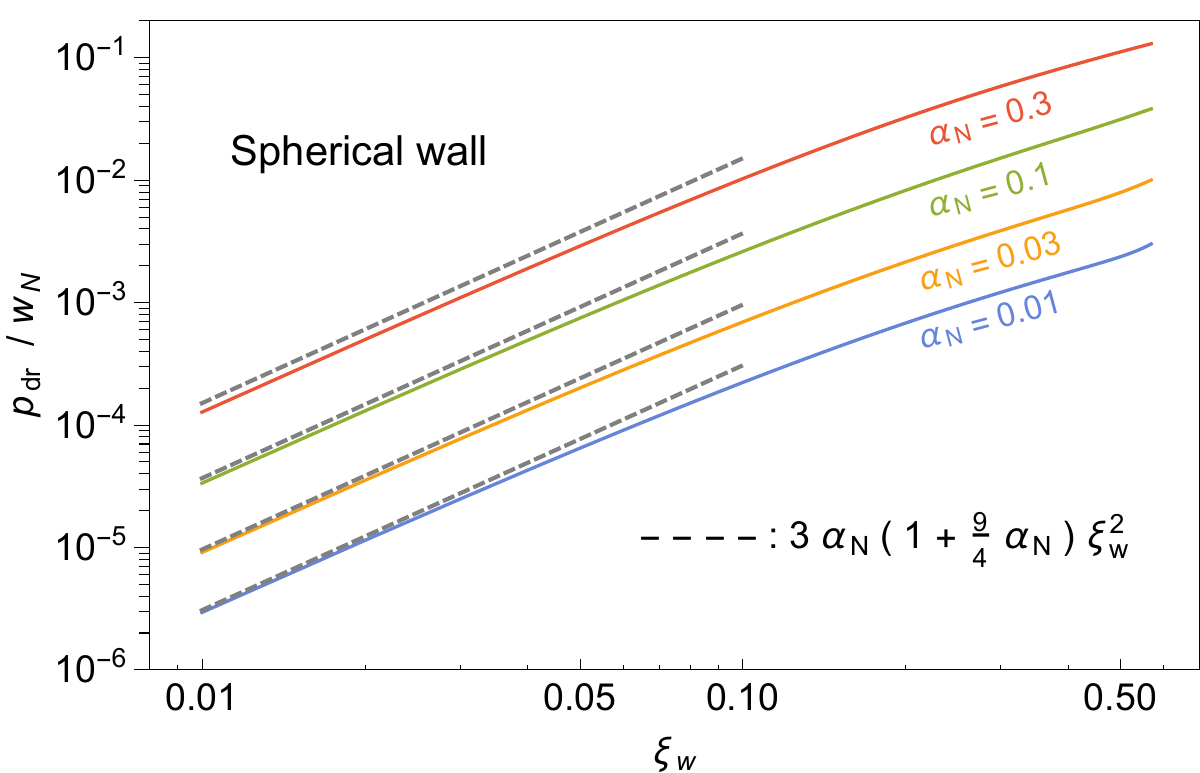}\\
\caption{Exact numerical evaluations for the phase pressure difference (rescaled by the asymptotic enthalpy) between false and true vacua as a function of the terminal velocity $\xi_w$ of planar (top), cylindrical (middle), and spherical (bottom) walls in the deflagration regime for given values of the asymptotic strength factor $\alpha_N$. Our analytic approximations are shown with black dashed lines in the non-relativistic regime.}\label{fig:StrongCoupling}
\end{figure}

For a planar wall with $D=0$ in~\eqref{eq:fluidEOM1},
\begin{align}\label{eq:fluidEOMPlanar}
\left(\mu(\xi,v)^2-c_s^2\right)\frac{\mathrm{d}v}{\mathrm{d}\xi} = 0,
\end{align}
the solutions are either $\mathrm{d}v/\mathrm{d}\xi=0$, namely, $v=\mathrm{const}$, or $\mu(\xi,v) = c_s$, which would lead to $\xi>c_s$ for $v>0$ but with no deflagration regime. Hence, the only solution should be $v = \mathrm{const}. = v_+$ in the sound shell, and the corresponding enthalpy profile from~\eqref{eq:fluidEOM2} with $\mathrm{d}v/\mathrm{d}\xi=0$ also stays constant in the sound shell, $w_+=\mathrm{const}.=w_L$. This $w_L$ can be related to $w_R=w_N$ by the junction condition~\eqref{eq:junctionshock} via $\tilde{v}_R=\xi_{sh}$ and $\tilde{v}_L=\mu(\xi_{sh},v_{sh})$ from the fluid velocity $v_{sh}$ at the shock front $\xi_{sh}$ in the background plasma frame. To further determine $v_{sh}$ and $\xi_{sh}$, note that the constant velocity profile in the sound shell implies $v_{sh}=v_+=\mu(\xi_w,\bar{v}_+(\bar{v}_-,\alpha_+))$ with $\bar{v}_- = \xi_w$ for the deflagration expansion and $\bar{v}_+(\bar{v}_-,\alpha_+)$ given by the minus-sign branch of Eq.~\eqref{eq:vpmbar}. Thus, $v_{sh}$ can be expressed in terms of $\xi_w$ and $\alpha_+$ alone. Once $v_{sh}$ is determined,  $\xi_{sh}$ can be directly obtained from the shock front condition $\mu(\xi_{sh},v_{sh})\xi_{sh}=c_s^2$. Hence, $w_+/w_N=\alpha_N/\alpha_+$ can be derived in terms of $\xi_w$ and $\alpha_+$ alone, which can be expanded at the LO in $\xi_w$ as 
\begin{align}
\left(\frac{\alpha_N}{\alpha_+}\right)_\mathrm{LO} = 1 + \left( \frac{1+c_s^2}{c_s^3}\right) \alpha_+ \xi_w + \mathcal{O}(\xi_w^2).
\end{align}
Reversing the above relation to arrive at $\alpha_+(\xi_w,\alpha_N)=(\sqrt{c_s^6+4c_s^3(1+c_s^2)\alpha_N\xi_w}-c_s^3)/[2(1+c_s^2)\xi_w]$, and then plugging it into the minus-sign branch of Eq.~\eqref{eq:vpmbar}, $\bar{v}_+(\bar{v}_-\equiv\xi_w,\alpha_+(\xi_w,\alpha_N))$, we can further expand $v_+=\mu(\xi_w,\bar{v}_+(\xi_w,\alpha_N))$ into $v_+=(\alpha_N/c_s^2)\xi_w-[(1+c_s^2)/c_s^5]\alpha_N^2\xi_w^2+c_s^{-4}\alpha_N[1-\alpha_N(1-c_s^2)/c_s^2+\alpha_N^2(2c_s^4+3c_s^2+2)/c_s^4]\xi_w^3+\mathcal{O}(\xi_w^4)$, which finally yields the phase pressure difference~\eqref{eq:FdriveDeflagration} in the small $\xi_w$ limit as
\begin{align}\label{eq:FdriveapproxPl}
\frac{p_\mathrm{dr}}{w_N}=\frac{\alpha_N}{c_s}\xi_w-\frac{\alpha_N(1+\alpha_N)}{c_s^2}\xi_w^2+\mathcal{O}(\xi_w^3).
\end{align}
This analytic approximation perfectly matches the exact numerical evaluation as shown in the top panel of Fig.~\ref{fig:StrongCoupling}.
In particular, the LO term is linear in $\xi_w$, reproducing the  linear trend between the phase pressure difference and terminal velocity of a planar wall found in the holographic numerical simulation for a strongly coupled non-Abelian gauge theory with a FOPT~\cite{Bea:2021zsu}. Nevertheless, a qualitative comparison to Ref.~\cite{Bea:2021zsu} is not feasible since the sound velocity (or equivalently the EOS) and strength factor $\alpha_N$ cannot be fixed independently in  the holographic numerical simulations.

\subsection{Cylindrical wall}

For a cylindrical wall with $D=1$, the fluid EOM~\eqref{eq:fluidEOM1} can be solved to the LO in $v$ from
\begin{align}
\frac{\mathrm{d}v}{\mathrm{d}\xi}=\frac{c_s^2v}{\xi(\xi^2-c_s^2)}+\mathcal{O}(v^2)
\end{align}
as $v(\xi)=C\sqrt{c_s^2/\xi^2 - 1}$ with $C$ fixed by $v(\xi_w)=v_+$, resulting in the LO solution $v_\mathrm{LO}(\xi)=v_+[(c_s^2/\xi^2 - 1)/(c_s^2/\xi_w^2 - 1)]^{1/2}$ and its inverse solution $\xi_\mathrm{LO}(v)$. It is easy to see that the shock front where $v_\mathrm{LO}(\xi)$ drops to zero is now approximated at $\xi_{sh}=c_s$ with $w(c_s)=w_N$, from which we can integrate the fluid EOM~\eqref{eq:fluidEOM2} to evaluate $w_+$ at $\xi_w$ from a LO estimation for $(\mathrm{d}\ln w/\mathrm{d}\xi)_\mathrm{LO}$. To estimate the LO derivative term $(\mathrm{d}\ln w/\mathrm{d}\xi)_\mathrm{LO}$, we can first expand the right-hand-side of~\eqref{eq:fluidEOM2} to the LO in $v$ after inserting the LO solution $\xi_\mathrm{LO}(v)$, and then plug back the LO solution $v_\mathrm{LO}(\xi)$ to express the LO derivative term $(\mathrm{d}\ln w/\mathrm{d}\xi)_\mathrm{LO}$ as a function of $\xi$. Hence, the LO estimation of $(\alpha_N/\alpha_+)_\mathrm{LO}=(w_+/w_N)_\mathrm{LO}$ is now a function of $\xi_w$, $\alpha_+$, and $v_+=\mu(\xi_w,\bar{v}_+)$, which, after inserting $\bar{v}_+(\xi_w,\alpha_+)$ from the minus-sign branch of~\eqref{eq:vpmbar}, can be expanded in the small $\xi_w$ limit as
\begin{align}\label{eq:alphaNapproxCy}
\left(\frac{\alpha_N}{\alpha_+}\right)_\mathrm{LO}=1 + \left(\frac{1+c_s^2}{c_s^4}\right)\alpha_+\xi_w^2 \ln\frac{2c_s}{\xi_w} + \mathcal{O}(\xi_w^4).
\end{align}
Reversing the above relation to get $\alpha_+(\xi_w,\alpha_N)$, and then plugging it into the minus-sign branch of Eq.~\eqref{eq:vpmbar}, $\bar{v}_+(\bar{v}_-=\xi_w,\alpha_+(\xi_w,\alpha_N))\equiv\bar{v}_+(\xi_w,\alpha_N)$, we can further expand $v_+=\mu(\xi_w,\bar{v}_+(\xi_w,\alpha_N))$ first in the small $\alpha_N$ limit and then in the small $\xi_w$ limit as $v_+=(\alpha_N/c_s^2)\xi_w+c_s^{-4}\alpha_N(1+\alpha_N-\alpha_N/c_s^2)\xi_w^3+[(1+c_s^2)/c_s^6]\alpha_N^2\xi_w^3\ln(\xi_w/2c_s)+\mathcal{O}(\xi_w^4)$. Finally, the phase pressure difference~\eqref{eq:FdriveDeflagration} is obtained in the small $\xi_w$ limit as
\begin{align}\label{eq:FdriveapproxCy}
\frac{p_{\mathrm{dr}}}{w_N} = \left[\frac{\alpha_N^2}{c_s^4+c_s^6}-\frac{\alpha_N}{c_s^2}\left(1+\ln\frac{\xi_w}{2c_s}\right)\right]\xi_w^2 + \mathcal{O}(\xi_w^4).
\end{align}
This analytic approximation also perfectly matches the exact numerical evaluation as shown in the middle panel of Fig.~\ref{fig:StrongCoupling}, of which the non-relativistic limit with a logarithmic dependence can be directly tested in the holographic numerical simulation of a strongly coupled FOPT with a cylindrical wall~\cite{Bea:2022mfb}.

\subsection{Spherical wall}

For a spherical wall with $D=2$, the fluid EOM~\eqref{eq:fluidEOM1} can be solved to the LO in $v$ from
\begin{align}
\frac{\mathrm{d}v}{\mathrm{d}\xi}=\frac{2c_s^2v}{\xi(\xi^2-c_s^2)}+\mathcal{O}(v^2)
\end{align}
as $v(\xi)=C(c_s^2/\xi^2-1)$ with $C$ fixed by $v(\xi_w)=v_+$, resulting in the LO solution $v_\mathrm{LO}(\xi)=v_+(c_s^2/\xi^2-1)/(c_s^2/\xi_w^2-1)$ and its inverse solution $\xi_\mathrm{LO}(v)$. Following the same procedures as in the cylindrical case, we can first obtain the LO derivative term $(\mathrm{d}\ln w/\mathrm{d}\xi)_\mathrm{LO}$, from which the enthalpy $w_+$ is integrated by the fluid EOM~\eqref{eq:fluidEOM2}, leading to a LO estimation of $(\alpha_N/\alpha_+)_\mathrm{LO}$ in the small $\xi_w$ limit as
\begin{align}\label{eq:alphaNapprox}
\left(\frac{\alpha_N}{\alpha_+}\right)_\mathrm{LO}=1+2\left(\frac{1+c_s^2}{c_s^4}\right)\alpha_+\xi_w^2+\mathcal{O}(\xi_w^3).
\end{align}
Reversing the above relation to get $\alpha_+(\xi_w,\alpha_N)$, the fluid velocity $v_+=\mu(\xi_w,\bar{v}_+(\xi_w,\alpha_+(\xi_w,\alpha_N)))$ can be expressed in terms of $\xi_w$ and $\alpha_N$ alone, which can be further expanded first in the small $\alpha_N$ limit and then in the small $\xi_w$ limit as $v_+=(\alpha_N/c_s^2)\xi_w+c_s^{-4}\alpha_N[1-\alpha_N(3+c_s^2)/c_s^2]\xi_w^3+\mathcal{O}(\xi_w^4)$. Finally, the phase pressure difference can be expanded in the small $\xi_w$ limit as
\begin{align}\label{eq:Fdriveapprox}
\frac{p_\mathrm{dr}}{w_N}=\alpha_N\left(\frac{1}{c_s^2}+\frac{\alpha_N}{c_s^4+c_s^6}\right)\xi_w^2+\mathcal{O}(\xi_w^4).
\end{align}
This analytic approximation again perfectly matches with the exact numerical evaluation for a small $\alpha_N$ but slightly deviates when $\alpha_N$ is larger as shown in the bottom panel of Fig.~\ref{fig:StrongCoupling}, of which the non-relativistic limit with a quadratic dependence can be directly tested in future holographic numerical simulations of a strongly coupled FOPT with a spherical wall~\cite{Cai:2022omk,He:2022amv}.

\section{Conclusions and discussions}

Numerical simulations using perfect fluid hydrodynamics could fail in the relativistic regime due to large perturbations (see, however, \cite{Lewicki:2022nba} for a particle-based simulation), and the microphysics approach solving full Boltzmann equations might also fail for the strongly coupled theory due to the lack of knowledge on strong dynamics. This is where comes the holographic approach to numerically simulate the bubble dynamics for a strongly coupled theory with a FOPT, and has recently unfolded a novel linear relation between the phase pressure difference and planar wall velocity in the non-relativistic limit, allowing for determining the wall velocity purely from the perfect fluid hydrodynamics and thus in terms of the EOS.

Using only perfect fluid hydrodynamics with a bag EOS in the non-relativistic limit, we have analytically reproduced the linear relation between the phase pressure difference away from the wall, $(p_--p_N)/w_N$, and the terminal planar wall velocity. We further predict more realistic cases with cylindrical and spherical walls, which can be directly tested in future numerical simulations of a strongly coupled theory with a FOPT. Once confirmed, the terminal spherical wall velocity for a strongly coupled FOPT can be directly estimated from~\eqref{eq:Fdriveapprox} as
\begin{align}
\xi_w&=\sqrt{\frac{c_s^4\delta}{\alpha_N+c_s^2+c_s^4}},\\
\delta=\frac{\Delta V_\mathrm{eff}}{\Delta V_0}&=\left(1-\frac{c_s^2}{1+c_s^2}\frac{\partial\ln\Delta V_\mathrm{eff}}{\partial\ln T}\right)^{-1}
\end{align}
without reference to the underlying microphysics in details. Besides the pressure difference away from the wall as obtained in the main text, we also compute the pressure difference just near the wall in the supplemental appendix, and find a universal scaling,  
\begin{align}
\frac{p_+-p_-}{w_N}=\frac{\alpha_N}{c_s^2}\left(1-\frac{\alpha_N}{c_s^2}\right)\xi_w^2+\mathcal{O}(\xi_w^3),
\end{align}
for all wall geometries with planar, cylindrical, and spherical symmetries. This university can be understood as the pressure difference across the wall does not care about the sound-shell profile that explicitly depends on the wall geometry. Other possible generalizations of our analytic approach could devote to improving the hydrodynamics by going beyond the bag EOS with a constant~\cite{Giese:2020znk,Giese:2020rtr,Wang:2020nzm,Wang:2023jto} or even varying~\cite{Wang:2022lyd} sound velocity and including shear and bulk viscosity as well as relaxing the thin-wall approximation.

\begin{acknowledgments}
We thank Mikel Sanchez-Garitaonandia and Matti Jarvinen for helpful correspondences.
This work is supported by the National Key Research and Development Program of China Grant  No. 2021YFC2203004, No. 2020YFC2201501, and No. 2021YFA0718304, 
the National Natural Science Foundation of China Grants No. 12105344, No. 12122513, No. 12075298, No. 11991052, No. 12235019, and No. 12047503, and by the Chinese Academy of Sciences Project for Young Scientists in Basic Research YSBR-006.
\end{acknowledgments}

\appendix

\section{Pressure difference on the wall}

In this supplemental appendix, we will analytically derive the pressure difference, $\Delta_\mathrm{wall}p\equiv p(\xi_w^+)-p(\xi_w^-)\equiv p_+-p_-$, just right between the front and back of the wall,  $\xi_w\pm(\epsilon\to0^+)\equiv\xi_w^\pm$, respectively. This is different from the phase  pressure difference we obtain in the main text, where the pressure difference is taken between the back of the bubble wall $\xi_w^-$ and the front of the shock front $\xi_{sh}+(\epsilon\to0^+)\equiv\xi_{sh}^+$, that is $p_\mathrm{dr}=\Delta V_\mathrm{eff}=-\Delta p=p(\xi=0)-p(\xi=1)=p(\xi_w^-)-p(\xi_{sh}^+)$.

However, it is usually difficult to extract sensible pressure difference purely on the wall as the perfect fluid hydrodynamics could fail around the bubble wall due to the temperature saltation and non-equilibrium effects across the bubble wall as also observed in the recent holographic numerical simulation~\cite{Bea:2021zsu}. Nevertheless, we have managed to separate out the wall contribution to the total pressure change between the false and true vacua in our previous study~\cite{Wang:2022txy} without assuming a bag EOS or local thermal equilibrium across the bubble wall.

In this supplemental appendix, we will first give a concise summary of the analytic results~\cite{Wang:2022txy} previously checked numerically with a bag EOS, and then provide here alternative analytic proof. Finally, we analytically derive a universal scaling for the pressure difference acting on the wall in the non-relativistic limit for planar, cylindrical, and spherical walls.

\subsection{Hydrodynamic backreaction force}

In our previous study~\cite{Wang:2022txy} for a steady-state thin-wall expansion with planar ($D=0$), cylindrical ($D=1$), and spherical ($D=2$) symmetries in the flat spacetime, we have proposed a hydrodynamic evaluation on the total backreaction force (per unit area)
\begin{align}\label{eq:Fbackhydro}
\frac{F_\mathrm{back}^\mathrm{hydro}}{A}&=
\int \mathrm{d}\xi \frac{t\nabla_\mu (w u^\mu)}{\gamma(\xi - v)}\nonumber\\
&=\int_0^1\mathrm{d}\xi\left(-\frac{\mathrm{d}w}{\mathrm{d}\xi}+\frac{Dwv}{\xi(\xi-v)}+\frac{w\gamma^2}{\mu}\frac{\mathrm{d}v}{\mathrm{d}\xi}\right)
\end{align}
that balances the driving force 
\begin{align}\label{eq:Fdrive}
\frac{F_\mathrm{drive}}{A}=\Delta V_\mathrm{eff}\equiv V_\mathrm{eff}(\phi_+)-V_\mathrm{eff}(\phi_-)
\end{align}
from the difference of the effective potential between false and true vacua. Here $\xi\equiv r/t$ is the self-similar coordinate tracing the fluid element at $r=\xi t$ after time lapse $t$ of bubble nucleation (ignoring the initial size)  so that $\xi=0$ and $\xi=1$ correspond to the bubble center and null infinity, respectively. In the background plasma frame, $v(\xi)$ and $w(\xi)$ are the fluid velocity and enthalpy profiles, respectively, and the abbreviation $\mu(\xi,v)\equiv(\xi-v)/(1-\xi v)$ is the Lorentz boost of the fluid velocity in a local frame moving with velocity $\xi$, and $\gamma(v)\equiv1/\sqrt{1-v^2}$ is an abbreviation for the Lorentz factor.  It is worth noting that our hydrodynamic expression~\eqref{eq:Fbackhydro} for the total backreaction force consists of the thermal force and friction force due to the temperature jumping and out-of-equilibrium effects across the bubble wall, respectively, and hence it is derived without assuming either a constant temperature profile or local thermal equilibrium across the wall interface.

It was found in Ref.~\cite{Wang:2022txy} that the hydrodynamic expression~\eqref{eq:Fbackhydro} for the total backreaction force can be further decomposed into the bubble-wall contribution, sound-shell contribution, and shock-front contribution (if any) as
\begin{align}
\frac{F_\mathrm{back}^\mathrm{hydro}}{A}\bigg|_\mathrm{wall}&=-\Delta_\mathrm{wall}w+\int_{v_-}^{v_+}\mathrm{d}v\frac{w(v)\gamma(v)^2}{\mu(\xi_w,v)},\label{eq:Fbackhydrowall}\\
\frac{F_\mathrm{back}^\mathrm{hydro}}{A}\bigg|_\mathrm{shell}&=-\int_\mathrm{shell}\mathrm{d}\xi\frac{\mathrm{d}w}{\mathrm{d}\xi}\frac{c_s^2}{1+c_s^2},\label{eq:Fbackhydroshell}\\
\frac{F_\mathrm{back}^\mathrm{hydro}}{A}\bigg|_\mathrm{shock}&=-\Delta_\mathrm{shock}w+\int_{v_{sh}}^{0}\mathrm{d}v\frac{w(v)\gamma(v)^2}{\mu(\xi_{sh},v)}\label{eq:Fbackhydroshock},
\end{align}
where $\xi_w$ and $\xi_{sh}$ are the velocities of the bubble wall and shock front, respectively, while $v_+$  and $v_-$ are the fluid velocities just in the front and back of the bubble wall, respectively, and $v_{sh}$ is the fluid velocity just in the back of the shockwave front (note here that we have used slightly different symbol from Ref.~\cite{Wang:2022txy}). The second integral~\eqref{eq:Fbackhydroshell} over the sound shell part has excluded the bubble-wall and shock-front contribution, and we have introduced the abbreviations
$\Delta_\mathrm{wall}w\equiv\lim_{\delta\to0^+}[w(\xi_w+\delta)-w(\xi_w-\delta)]\equiv w_+-w_-$ and
$\Delta_\mathrm{shock}w\equiv\lim_{\delta\to0^+}[w(\xi_{sh}+\delta)-w(\xi_{sh}-\delta)]\equiv w_R-w_L$
for the enthalpy differences across the bubble wall and shock front, respectively.

As shown numerically in the previous study \cite{Wang:2022txy}, the wall contribution~\eqref{eq:Fbackhydrowall} can reproduce a previous estimation~\cite{BarrosoMancha:2020fay} for the backreaction force acting on the wall interface from the junction condition of the energy-momentum tensor, $w_-\bar{v}_-^2\bar{\gamma}_-^2+p_-=w_+\bar{v}_+^2\bar{\gamma}_+^2+p_+$, that is 
\begin{align}\label{eq:Fbackwall}
\frac{F_\mathrm{back}^\mathrm{hydro}}{A}\bigg|_\mathrm{wall}=\bar{\gamma}_+^2\bar{v}_+^2w_+-\bar{\gamma}_-^2\bar{v}_-^2w_-\equiv\Delta_\mathrm{wall}(\bar{\gamma}^2\bar{v}^2w),
\end{align}
where $\bar{v}_\pm=\mu(\xi_w,v_\pm)$ is the fluid velocity $v_\pm$ but seen by an observer comoving with the bubble wall, and $\bar{\gamma}_\pm\equiv\gamma(\bar{v}_\pm)$ are the corresponding Lorentz factors. In this appendix, we will further prove analytically this equivalence~\eqref{eq:Fbackwall} as well as the balance of our hydrodynamic evaluation on the total backreaction force~\eqref{eq:Fbackhydro} against the driving force~\eqref{eq:Fdrive},
\begin{align}\label{eq:FdriveBagEoS}
\frac{F_\mathrm{drive}}{A}=-\frac14\Delta w+\frac34\alpha_Nw_N,
\end{align}
for a bag equation of state with the sound velocity $c_s=1/\sqrt{3}$, where the enthalpy difference $\Delta w\equiv w_N-w_O$ is taken between the null infinity $w_N\equiv w(\xi=1)$ and bubble center $w_O\equiv w(\xi=0)$, and $\alpha_N\equiv4\Delta V_0/(3w_N)$ is the strength factor at the null infinity determined by the difference in the vacuum potential energy density $\Delta V_0$ and the enthalpy $w_N$ at the null infinity.  The proof goes as follows for three different modes of fluid motions.

\subsection{Alternative analytic proof}

The main trick for the analytic proof below is to contribute the  balance of forces to the physical branch of hydrodynamic solutions~\cite{Espinosa:2010hh},
\begin{align}\label{eq:vpmbarapp}
\bar{v}_+=\frac{1}{1+\alpha_+}\left(X_+\pm\sqrt{X_-^2+\alpha_+^2+\frac23\alpha_+}\right)
\end{align}
with abbreviations $X_\pm\equiv\bar{v}_-/2\pm1/(6\bar{v}_-)$ and $\alpha_+\equiv4\Delta V_0/(3w_+)$, where the  detonation expansion ($1/\sqrt{3}\equiv c_s<\bar{v}_-<\bar{v}_+\equiv\xi_w$) picks the plus-sign branch of \eqref{eq:vpmbarapp}, while the deflagration expansion ($\bar{v}_+<\bar{v}_-=\xi_w<c_s\equiv1/\sqrt{3}$) and hybrid expansion ($\bar{v}_+<\bar{v}_-=c_s\equiv1/\sqrt{3}$) picks the minus-sign branch of \eqref{eq:vpmbarapp}.

\subsubsection{Detonation}

For the detonation mode, the backreaction force reads
\begin{align}\label{eq:FbackhydroDetona}
\frac{F_\mathrm{back}^\mathrm{hydro}}{A}=\frac{F_\mathrm{back}^\mathrm{hydro}}{A}\bigg|_\mathrm{wall}+\frac{F_\mathrm{back}^\mathrm{hydro}}{A}\bigg|_\mathrm{shell}
\end{align}
with
\begin{align}
\frac{F_\mathrm{back}^\mathrm{hydro}}{A}\bigg|_\mathrm{wall}&=-(w_+-w_-)+\frac{v_-}{v_--\xi_w}w_+,\\
\frac{F_\mathrm{back}^\mathrm{hydro}}{A}\bigg|_\mathrm{shell}&=-\frac14(w_--w_s),
\end{align}
where the wall contribution reproduces the equivalence
\begin{align}
\frac{F_\mathrm{back}^\mathrm{hydro}}{A}\bigg|_\mathrm{wall}=\frac{v_-\xi_w}{1-v_-\xi_w}w_+=\Delta_\mathrm{wall}(\bar{\gamma}^2\bar{v}^2w)
\end{align}
after inserting the junction condition
\begin{align}
w_-=\frac{\bar{\gamma}_+^2\bar{v}_+}{\bar{\gamma}_-^2\bar{v}_-}w_+=\frac{\xi_w}{1-\xi_w^2}\frac{1-\mu(\xi_w,v_-)^2}{\mu(\xi_w,v_-)}w_+
\end{align}
and the replacements $\bar{v}_+=\xi_w$ and $\bar{v}_-=\mu(\xi_w,v_-)$.
Note that the driving force can be rearranged as
\begin{align}
\frac{F_\mathrm{drive}}{A}&=-\frac14(w_N-w_O)+\frac34\alpha_Nw_N,\nonumber\\
&=-\frac14[(w_+-w_-)+(w_--w_s)]+\frac34\alpha_+w_+
\end{align}
with $w_O=w(\xi=c_s)\equiv w_s$ and $w_N=w_+$ as well as the identity $\alpha_Nw_N=\alpha_+w_+\equiv4\Delta V_0/3$, therefore, the balance of the driving force against the total backreaction force~\eqref{eq:FbackhydroDetona} wound render another equivalence
\begin{align}
\frac{F_\mathrm{back}^\mathrm{hydro}}{A}\bigg|_\mathrm{wall}=-\frac14(w_+-w_-)+\frac34\alpha_+w_+,
\end{align}
leading to a relation
\begin{align}
\alpha_+=\frac{v_-(3\xi_w^2-2v_-\xi_w-1)}{3(\xi_w-v_-)(1-v_-\xi_w)},
\end{align}
which is nothing but the plus-sign branch of~\eqref{eq:vpmbarapp} with $\bar{v}_+=\xi_w$. As a result, the balance between the driving force and backreaction force is also analytically proved for the detonation mode. 

\subsubsection{Deflagration}

For the deflagration mode, the backreaction force is given by
\begin{align}\label{eq:FbackhydroDeflag}
\frac{F_\mathrm{back}^\mathrm{hydro}}{A}=\frac{F_\mathrm{back}^\mathrm{hydro}}{A}\bigg|_\mathrm{wall}+\frac{F_\mathrm{back}^\mathrm{hydro}}{A}\bigg|_\mathrm{shell}+\frac{F_\mathrm{back}^\mathrm{hydro}}{A}\bigg|_\mathrm{shock}
\end{align}
with
\begin{align}
\frac{F_\mathrm{back}^\mathrm{hydro}}{A}\bigg|_\mathrm{wall}&=-(w_+-w_-)-\frac{v_+}{v_+-\xi_w}w_-,\\
\frac{F_\mathrm{back}^\mathrm{hydro}}{A}\bigg|_\mathrm{shell}&=-\frac14(w_L-w_+),\\
\frac{F_\mathrm{back}^\mathrm{hydro}}{A}\bigg|_\mathrm{shock}&=-(w_R-w_L)+\frac{v_{sh}}{v_{sh}-\xi_{sh}}w_R.\label{eq:FbackhydroshockDeflag}
\end{align}
It is straightforward to check the equivalence
\begin{align}
\frac{F_\mathrm{back}^\mathrm{hydro}}{A}\bigg|_\mathrm{wall}=\frac{v_+^2-v_+\xi_w}{1-v_+^2}w_+=\Delta_\mathrm{wall}(\bar{\gamma}^2\bar{v}^2w)
\end{align}
after inserting the junction condition
\begin{align}
w_-=\frac{\bar{\gamma}_+^2\bar{v}_+}{\bar{\gamma}_-^2\bar{v}_-}w_+=\frac{\bar{v}_+}{1-\bar{v}_+^2}\frac{1-\xi_w^2}{\xi_w}w_+
\end{align}
and the replacements $\bar{v}_+=\mu(\xi_w,v_+)$ and $\bar{v}_-=\xi_w$. For the shock contribution of the backreaction force~\eqref{eq:FbackhydroshockDeflag}, it is easy to check that
\begin{align}
\frac{F_\mathrm{back}^\mathrm{hydro}}{A}\bigg|_\mathrm{shock}=\frac{3\xi_{sh}^2-1}{3(1-\xi_{sh}^2)}w_R=-\frac14(w_R-w_L)
\end{align}
after inserting the junction condition
\begin{align}\label{eq:JunctionDeflag}
w_L=\frac{\tilde{\gamma}_R^2\tilde{v}_R}{\tilde{\gamma}_L^2\tilde{v}_L}w_R = \frac{\xi_{sh}}{1-\xi_{sh}^2}\frac{1-\mu(\xi_{sh},v_{sh})^2}{\mu(\xi_{sh},v_{sh})}w_R
\end{align}
and the replacements $\tilde{v}_R=\xi_{sh}$ and $\tilde{v}_L=\mu(\xi_{sh},v_{sh})$. Note that the driving force can be rearranged as
\begin{align}
\frac{F_\mathrm{drive}^\mathrm{hydro}}{A}
=&-\frac14(w_N-w_O)+\frac34\alpha_Nw_N\nonumber\\
=&-\frac14[(w_R-w_L)+(w_L-w_+)+(w_+-w_-)]\nonumber\\
&+\frac34\alpha_+w_+
\end{align}
with $w_O=w_-$ and $w_N=w_R$ as well as the identity $\alpha_Nw_N=\alpha_+w_+\equiv4\Delta V_0/3$, therefore, the balance of the driving force against the total backreaction force~\eqref{eq:FbackhydroDeflag} would render another equivalence
\begin{align}
\frac{F_\mathrm{back}^\mathrm{hydro}}{A}\bigg|_\mathrm{wall}=-\frac14(w_+-w_-)+\frac34\alpha_+w_+,
\end{align}
leading to a relation
\begin{align}\label{eq:DeflagrationRelation}
\alpha_+=\frac{v_+(1+2v_+\xi_w-3\xi_w^2)}{3(1-v_+^2)\xi_w},
\end{align}
which is nothing but the minus-sign branch of~\eqref{eq:vpmbarapp} with $\bar{v}_-=\xi_w$. As a result, the balance between the driving force and backreaction force is also analytically proved for the deflagration mode.

\subsubsection{Hybrid}

For the hybrid mode, the backreaction force reads
\begin{align}\label{eq:FbackhydroHybrid}
\frac{F_\mathrm{back}^\mathrm{hydro}}{A}=\frac{F_\mathrm{back}^\mathrm{hydro}}{A}\bigg|_\mathrm{wall}+\frac{F_\mathrm{back}^\mathrm{hydro}}{A}\bigg|_\mathrm{shell}+\frac{F_\mathrm{back}^\mathrm{hydro}}{A}\bigg|_\mathrm{shock}
\end{align}
with
\begin{align}
\frac{F_\mathrm{back}^\mathrm{hydro}}{A}\bigg|_\mathrm{wall}&=-(w_+-w_-)+\frac{c_s(1-\xi_w^2)(v_--v_+)w_-}{(1-c_s^2)(v_--\xi_w)(\xi_w-v_+)},\\
\frac{F_\mathrm{back}^\mathrm{hydro}}{A}\bigg|_\mathrm{shell}&=-\frac14[(w_L-w_+)+(w_--w_s)],\\
\frac{F_\mathrm{back}^\mathrm{hydro}}{A}\bigg|_\mathrm{shock}&=-(w_R-w_L)+\frac{v_{sh}}{v_{sh}-\xi_{sh}}w_R.\label{eq:FbackhydroshockHybrid}
\end{align}
It is straightforward to check the equivalence
\begin{align}
\frac{F_\mathrm{back}^\mathrm{hydro}}{A}\bigg|_\mathrm{wall}=\frac{(\bar{v}_+-c_s)\bar{v}_+}{1-\bar{v}_+^2}w_+=\Delta_\mathrm{wall}(\bar{\gamma}^2\bar{v}^2w)
\end{align}
after inserting the junction condition
\begin{align}
w_-=\frac{\bar{\gamma}_+^2\bar{v}_+}{\bar{\gamma}_-^2\bar{v}_-}w_+=\frac{\bar{v}_+}{1-\bar{v}_+^2}\frac{1-c_s^2}{c_s}w_+
\end{align}
and the replacements $v_+=\mu(\xi_w,\bar{v}_+)$ and $v_-=\mu(\xi_w,c_s)$. For the shock contribution of the backreaction force~\eqref{eq:FbackhydroshockHybrid}, it is easy to check that
\begin{align}
\frac{F_\mathrm{back}^\mathrm{hydro}}{A}\bigg|_\mathrm{shock}=\frac{3\xi_{sh}^2-1}{3(1-\xi_{sh}^2)}w_R=-\frac14(w_R-w_L)
\end{align}
after inserting the junction condition
\begin{align}
w_L=\frac{\tilde{\gamma}_R^2\tilde{v}_R}{\tilde{\gamma}_L^2\tilde{v}_L}w_R=\frac{\xi_{sh}}{1-\xi_{sh}^2}\frac{1-\mu(\xi_{sh},v_{sh})^2}{\mu(\xi_{sh},v_{sh})}w_R
\end{align}
and the replacements $\tilde{v}_R=\xi_{sh}$ and $\tilde{v}_L=\mu(\xi_{sh},v_{sh})$. Note that the driving force can be rearranged as
\begin{align}
\frac{F_\mathrm{drive}^\mathrm{hydro}}{A}
=&-\frac14(w_N-w_O)+\frac34\alpha_Nw_N\nonumber\\
=&-\frac14[(w_R-w_L)+(w_L-w_+)+(w_--w_s)]\nonumber\\
&-\frac14(w_+-w_-)+\frac34\alpha_+w_+
\end{align}
with $w_O=w_s$ and $w_N=w_R$ as well as the identity $\alpha_Nw_N=\alpha_+w_+\equiv4\Delta V_0/3$, therefore, the balance of the driving force against the total backreaction force~\eqref{eq:FbackhydroHybrid} would render another equivalence
\begin{align}
\frac{F_\mathrm{back}^\mathrm{hydro}}{A}\bigg|_\mathrm{wall}=-\frac14(w_+-w_-)+\frac34\alpha_+w_+,
\end{align}
leading to a relation
\begin{align}
\alpha_+=\frac{(\bar{v}_+-c_s)(3c_s\bar{v}_+-1)}{3c_s(1-\bar{v}_+^2)},
\end{align}
which is nothing but the minus-sign branch of~\eqref{eq:vpmbarapp} with $\bar{v}_-=c_s$. As a result, the balance between the driving force and backreaction force is also analytically proved for the hybrid mode.

\subsection{Non-relativistic limit}

\begin{figure}
\begin{flushleft}
\includegraphics[width=0.45\textwidth]{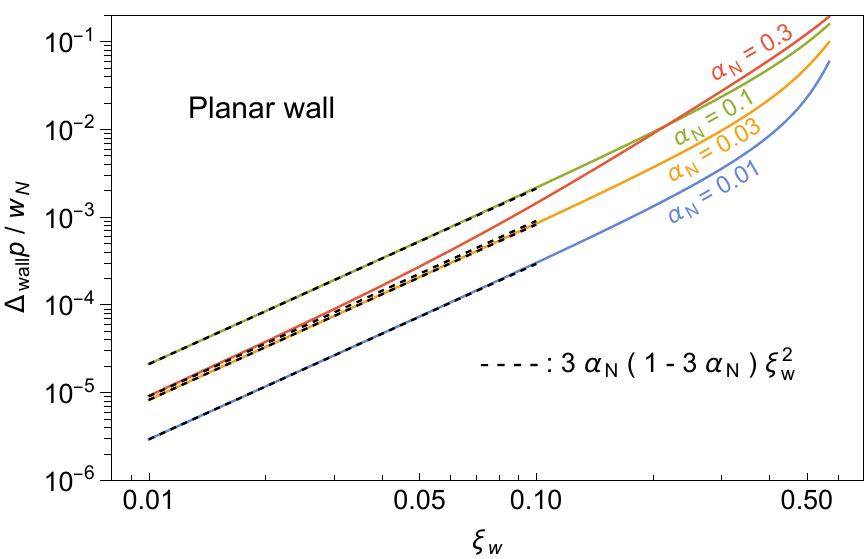}\\
\includegraphics[width=0.45\textwidth]{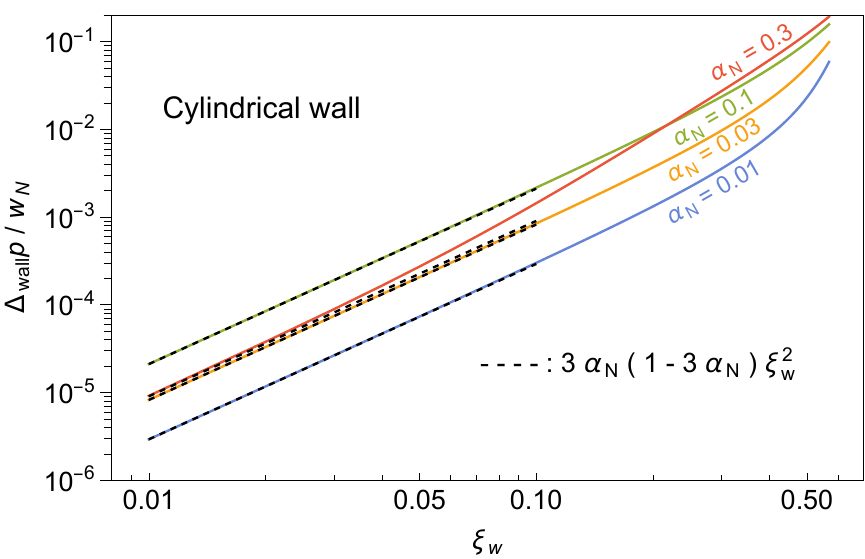}\\
\includegraphics[width=0.45\textwidth]{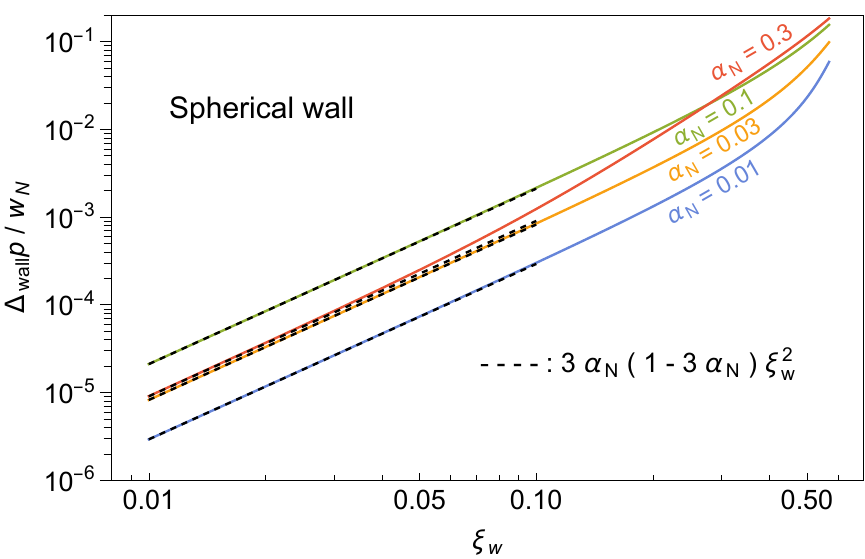}\\
\end{flushleft}
\caption{The pressure difference (rescaled by the asymptotic enthalpy) between the front and back of the wall as a function of the terminal velocity of planar (top), cylindrical (middle), and spherical (bottom) walls in the deflagration regime for given values of the asymptotic strength factor $\alpha_N$. Our analytic approximations are shown with black dashed lines in the non-relativistic regime.}\label{fig:PressureWall}
\end{figure}

For the deflagration expansion, the wall contribution of the total backreactioin force, $(F_\mathrm{back}^\mathrm{hydro}/Aw_N)_\mathrm{wall}=\Delta_\mathrm{wall}(\bar{\gamma}^2\bar{v}^2w)=-\Delta_\mathrm{wall}p$ is negative as confirmed by numerical calculations, therefore, the pressure difference taken just right in front and back of the wall scales as
\begin{align}
\frac{\Delta_\mathrm{wall}p}{w_N}&=\frac{w_+-w_-}{w_N}+\frac{v_+}{v_+-\xi_w}\frac{w_-}{w_N}\\
&=\frac{3\xi_w(\xi_w-v_+)\alpha_N}{1+2v_+\xi_w-3\xi_w^2},
\end{align}
where we have first converted $w_-$ to $w_+$ via the junction condition~\eqref{eq:JunctionDeflag} with $\bar{v}_+=\mu(\xi_w,v_+)$, and then converted $w_+$ to $w_N$ via the relation~\eqref{eq:DeflagrationRelation} to arrive at the second line. Finally, plugging in the non-relativistic analytic approximations $v_+(\xi_w,\alpha_N)$ we obtain in the main text for planar, cylindrical, and spherical walls,
\begin{align}
D=0: v_+&=3\alpha_N\xi_w -12\sqrt{3}\alpha_N^2 \xi_w^2\nonumber\\
&\quad+9\alpha_N(1-2\alpha_N + 29\alpha_N^2)\xi_w^3+\mathcal{O}(\xi_w^4),\\
D=1: v_+&=3\alpha_N\xi_w+9\alpha_N(1 - 2\alpha_N)\xi_w^3 \nonumber\\ 
&\quad+18\alpha_N^2\xi_w^3\ln\left(\frac{3}{4}\xi_w^2\right) + \mathcal{O}(\xi_w^4),\\
D=2: v_+&=3\alpha_N\xi_w+9\alpha_N(1-10\alpha_N)\xi_w^3\nonumber\\
&\quad+\mathcal{O}(\xi_w^4),
\end{align}
we expand the pressure difference near the wall in $\xi_w$ as
\begin{align}
\left(\frac{p_+-p_-}{w_N}\right)_{D=0}&=3\alpha_N(1-3\alpha_N)\xi_w^2+36\sqrt{3}\alpha_N^3\xi_w^3\nonumber\\
&\quad+9\alpha_N(1-8\alpha_N+12\alpha_N^2-87\alpha_N^3)\xi_w^4\nonumber\\
&\quad+\mathcal{O}(\xi_w^5),\\
\left(\frac{p_+-p_-}{w_N}\right)_{D=1}&=3\alpha_N(1-3\alpha_N)\xi_w^2-54\alpha_N^3\xi_w^4\ln\left(\frac{3}{4}\xi_w^2\right)\nonumber\\
&\quad+9\alpha_N(1-8\alpha_N+12\alpha_N^2)\xi_w^4\nonumber\\
&\quad+\mathcal{O}(\xi_w^5),\\
\left(\frac{p_+-p_-}{w_N}\right)_{D=2}&=3\alpha_N(1-3\alpha_N)\xi_w^2\nonumber\\
&\quad+9\alpha_N(1-8\alpha_N+36\alpha_N^2)\xi_w^4\nonumber\\
&\quad+\mathcal{O}(\xi_w^5),
\end{align}
where the LO contributions are all the same for different wall geometries,
\begin{align}
\left(\frac{p_+-p_-}{w_N}\right)_\mathrm{LO}=3\alpha_N(1-3\alpha_N)\xi_w^2,
\end{align}
as shown with dashed lines  in Fig.~\ref{fig:PressureWall}. This universal scaling can be understood as the pressure difference taken near the wall does not care about its global shape, while the pressure difference taken away from the wall (including the sound-shell and shock-front contributions) as derived in the main text depends on the wall geometry since the included sound shell actually shapes differently for different wall geometries.


\bibliography{ref}

\end{document}